\documentclass{iopart}

\usepackage{iopams}
\usepackage{graphicx}
\usepackage{cite}
\begin{document}
\title[Stable bunch trains for PWFA]{Stable bunch trains for plasma wakefield acceleration}

\author{K.V. Lotov}

\address{Budker Institute of Nuclear Physics, Novosibirsk, 630090, Russia}
\address{Novosibirsk State University, Novosibirsk, 630090, Russia}
\ead{K.V.Lotov@inp.nsk.su}
\vspace{10pt}
\begin{indented}
\item[]\today
\end{indented}

\begin{abstract}
A train of short charged particle bunches can efficiently drive a strong plasma wakefield over a long propagation distance only if all bunches reside in focusing and decelerating phases of the wakefield. This is shown possible with equidistant bunch trains, but requires the bunch charge to increase along the train and the plasma frequency to be higher than the bunch repetition frequency.
\end{abstract}

%
\vspace{2pc}

\section{Introduction}

Acceleration of particles in plasmas is now studied as a possible path to the future
of high-energy particle physics \cite{NatPhot7-775,PPCF56-084013,UFN55-965,RAST9-1,RAST9-209,RAST9-63}. In this method, laser or particle drive beams excite strong plasma waves that accelerate electrons or positrons by electric fields orders of magnitude higher than those in conventional accelerating structures \cite{RAST9-19,RMP81-1229,IEEE-PS24-252}. Among presently available drivers, only high-energy proton beams produced by modern synchrotrons can bring accelerated particles to TeV-range energies in a single plasma stage \cite{NIMA-829-3,PoP18-103101,RAST9-85}. This generates interest in studies of long-term dynamics of proton beams in plasmas.

Proton beams from synchrotrons are tens of centimeters long and cannot excite the plasma wave as they are. The typical wavelength of the plasma wave is about 1\,mm or shorter, so the proton beam must be micro-bunched at this wavelength to provide efficient coupling to the wave. The bunching can be done either prior to the plasma section with conventional techniques \cite{RuPAC16-303}, or by the plasma itself through the seeded self-modulation (SSM) of the proton bunch \cite{PRL104-255003,PoP22-103110}. The latter technique was already demonstrated by AWAKE experiment at CERN \cite{AWAKE,NIMA-829-76,Patric}, while the conventional method still waits for a proof-of-principle demonstration.

SSM modulates the proton beam by defocusing unnecessary beam parts transversely, so it is charge inefficient. At best, one quarter of the initial beam charge remains after self-modulation \cite{PoP22-103110}. The conventional micro-bunching uses longitudinal re-distribution of beam particles, and the beam can preserve its total charge. Consequently, the conventional micro-bunching technique is preferable for future applications of proton-driven plasma wakefield acceleration.

The micro-bunched proton beam, or the bunch train, must propagate in the plasma for a long distance without changing shape. This is only possible if each bunch resides in the focusing phase of the plasma wave. The bunch train produced by SSM is stable and focused everywhere \cite{PoP22-103110,PoP22-123107}, if the self-modulation develops in a longitudinally nonuniform plasma with a ramped-up density \cite{PoP18-024501}. However, the equidistant bunch train produced with conventional techniques is not. The question thus appears how can we approach the all-focused beam state with an equidistant bunch train.

\begin{figure}[t]\centering
\includegraphics[width=200bp]{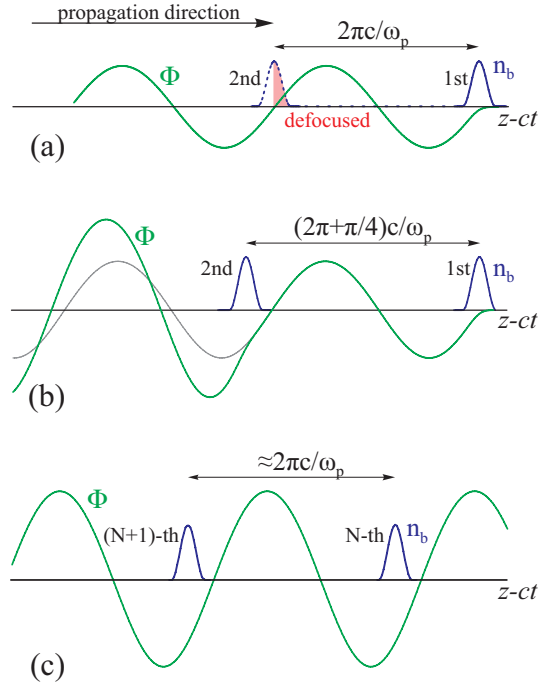}
\caption{Relative phasing of short proton bunches. The bunch density $n_b$ is shown in blue, and the wakefield potential $\Phi$ on the axis in green: (a) single bunch, (b) first two bunches (grey line shows the unloaded potential), (c) bunches in a grown wakefield.}\label{f1-distances}
\end{figure}

Figure~\ref{f1-distances}(a) illustrates the problem. Here we plot the wakefield potential $\Phi$ generated in the plasma by a short proton bunch that moves with the speed of light $c$. We assume the plasma responds linearly to the beam \cite{PAcc20-171}, since this is the case of interest for resonant wave excitation by multiple bunches \cite{PoP20-083119}. The wakefield potential contains information on both longitudinal $F_\parallel$ and transverse $F_\perp$ forces exerted by electric ($\vec{E}$) and magnetic ($\vec{B}$) fields on an axially moving speed-of-light proton:
\begin{equation}\label{e1}
    F_\parallel = e E_z = -e \frac{\partial \Phi}{\partial z}, \quad
    F_\perp = e (E_r - B_\phi) = -e \frac{\partial \Phi}{\partial r},
\end{equation}
where $e$ is the elementary charge, and, for simplicity, we assume the axial symmetry of the system with the $z$-axis being the direction of beam propagation and use cylindrical coordinates $(r,\phi,z)$. The plasma wave period in this case is exactly $2 \pi c/\omega_p$, where $\omega_p = \sqrt{4 \pi n e^2 / m}$ is the plasma frequency, $m$ is the electron mass, and $n$ is the plasma density. Regions of $\Phi < 0$ are potential wells in which protons are radially confined. If the second proton bunch follows the first one at exactly the plasma wave period, then its leading half is in a region of $\Phi > 0$ and will be defocused soon by the wakefield of the first bunch.

Mismatching the bunch frequency and the plasma frequency cannot solve the problem, since the required extension of the bunch-to-bunch distance depends on the bunch position in the train \cite{AIP396-75}. For example, the second bunch strongly modifies the wakefield of the first bunch and is focused if located  about $(2\pi + \pi/4)c/\omega_p$ behind the first bunch [Fig.\,\ref{f1-distances}(b)]. If the bunches are somewhere in the middle of the train, then they propagate in a well-grown wakefield and modify it by only a small amount, so the required bunch-to-bunch distance is close to $2\pi c/\omega_p$ [Fig.\,\ref{f1-distances}(c)].

However, all bunches of an equidistant bunch train can be in focusing and decelerating phases of the wave in the case of variable bunch charges. In this paper, we analyze this option and show that the required bunch charge distribution is rather exotic (exponential), but can be approximated by a bunched Gaussian beam.

\begin{figure}[t]\centering
\includegraphics[width=210bp]{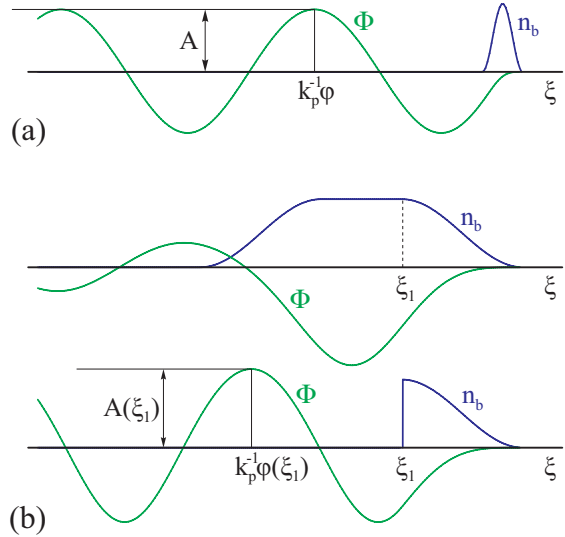}
\caption{Defining amplitude $A$ and phase $\phi$ of the wakefield potential: (a) directly between the bunches, and (b) with an auxiliary cut beam.}\label{f2-amplitudes}
\end{figure}

\section{Stable bunch train}

\begin{figure}[t]\centering
\includegraphics[width=195bp]{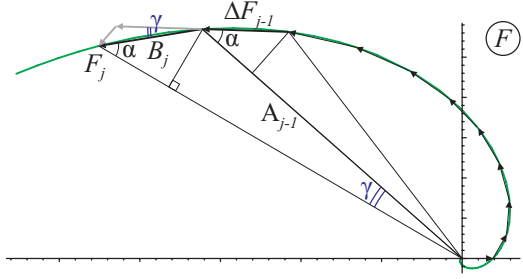}
\caption{Illustration of complex amplitude growth.}\label{f3-spiral}
\end{figure}

We attack the problem with the method of complex potential amplitudes developed in \cite{PoP22-103110}. Namely, we present the on-axis wakefield potential as
\begin{equation}\label{e2}
    \Phi (z,t) = {\rm{Re}} \left( A (\xi) e^{i [k_p \xi-\phi(\xi)]} \right),
\end{equation}
where $\xi = z-ct$ is the co-moving coordinate, $k_p = \omega_p/c$, and amplitude $A$ and phase $\phi$ are constant between the bunches [Fig.\,\ref{f2-amplitudes}(a)]. We assume the plasma responds linearly to the beam, and dissipation of the plasma wave is negligibly small. Inside the bunches, the amplitude and the phase at some $\xi=\xi_1$ can be determined by assuming zero beam density at $\xi<\xi_1$  [Fig.\,\ref{f2-amplitudes}(b)] and measuring the wakefield there. Real-valued amplitude $A$ and phase $\phi$ of the wakefield potential can then be combined into the complex amplitude
\begin{equation}\label{e3}
    F (\xi) = A e^{-i \phi}
\end{equation}
that contains information on both the amplitude and the phase of the wave. Each bunch in the train makes some contribution $\Delta F_j$ to the complex amplitude $F$ and leaves behind a wave of complex amplitude $F_j$, where $j$ is the bunch number (Fig.\,\ref{f3-spiral}). The charge and shape of the bunch determine the absolute value of this contribution $B_j = |\Delta F_j|$. The location of the bunch determines the argument of $\Delta F_j$. Consecutive contributions of individual bunches form a polyline on the complex plane $F$. Rotation of this polyline with respect to the origin does not change the essence of the process, so only angles between bunch contributions $\Delta F_j$ and wakefield states $F_j$ are of importance.

\begin{figure}[b]\centering
\includegraphics[width=210bp]{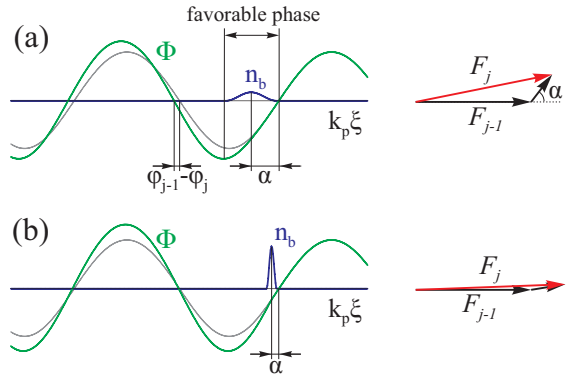}
\caption{Contributions to the wakefield potential of long (a) and short (b) bunches.}\label{f4-bunches}
\end{figure}

Optimally positioned bunches form a spiral on the complex plane $F$ \cite{PoP22-103110}, so that each bunch increases the wave amplitude and advances the wave phase (Fig.\,\ref{f3-spiral}). The word ``optimally'' means the bunch experiences the strongest decelerating field and, at the same time, is completely in the potential well. The spiral angle $\alpha$ equals $\pi/4$ for long bunches that cover full favorable region [Fig.\,\ref{f4-bunches}(a)] and approaches zero for short bunches [Fig.\,\ref{f4-bunches}(b)].

As we see from Fig.\,\ref{f3-spiral}, the contributions of individual bunches can fit into the spiral only if each subsequent contribution is rotated relative to the previous one by some angle $\gamma$. This angle appears if the repetition frequency of the bunches differs from the plasma frequency. The bunch-to-bunch distance is $1+\gamma/(2\pi)$ times the plasma wavelength, and the bunch repetition frequency
\begin{equation}\label{e4}
    \omega_b = \frac{\omega_p}{1+\gamma/(2\pi)}.
\end{equation}
Correspondingly, the plasma density must be higher than the density $n_{\rm res}$ providing the exact resonance with the bunch train:
\begin{equation}\label{e5}
   n = n_{\rm res} (1+\gamma/(2\pi))^2.
\end{equation}
A train of equal equidistant bunches cannot follow the optimum spiral. There is a choice between being equidistant ($\gamma=\rm const$) and being equal ($B_j=\rm const$). For equidistant bunches, the contributions of bunches must be different in absolute value.

\begin{figure}[tbh]\centering
\includegraphics[width=159bp]{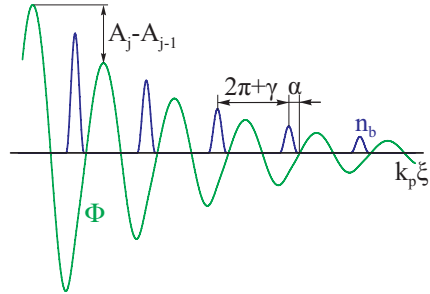}
\caption{An example of an all-focused equidistant bunch train.}\label{f5-ideal}
\end{figure}

Now we can use geometrical considerations to establish relationships between parameters of the optimum bunch train. We assume the train contains a large number of bunches, and the contribution of an individual bunch is small compared to the wakefield created by preceding bunches. Then the phase advance $\gamma$ produced by each bunch is small and related to the bunch contribution $B_j$ as
\begin{equation}\label{e6}
    \gamma = |\phi_{j} - \phi_{j-1}| \approx \frac{B_j \sin\alpha}{A_{j-1}}.
\end{equation}
The wakefield amplitude grows exponentially,
\begin{equation}\label{e7}
    A_j = A_{j-1} + B_j \cos \alpha \approx A_{j-1} (1 + \gamma \cot \alpha),
\end{equation}
\begin{equation}\label{e8}
    A_j \approx A_0 e^{j \gamma \cot \alpha}, \qquad F_j \approx F_0 e^{j \gamma (i + \cot \alpha)},
\end{equation}
as do contributions of individual bunches. Changing from the bunch number $j$ to the co-moving coordinate $\xi = -j k_p^{-1}(2\pi+\gamma)$ yields
\begin{equation}\label{e9}
    A \propto \exp\left( -\frac{k_p \xi \gamma \cot \alpha}{2\pi+\gamma}\right).
\end{equation}
Therefore, the required plasma density (through the parameter $\gamma$), the bunch shape (through the parameter $\alpha$), the wakefield growth rate, and the relative contributions of individual bunches are uniquely related to each other. Figure~\ref{f5-ideal} illustrates this optimum bunch train and the wakefield potential generated by it.

\begin{figure}[tbh]\centering
\includegraphics[width=171bp]{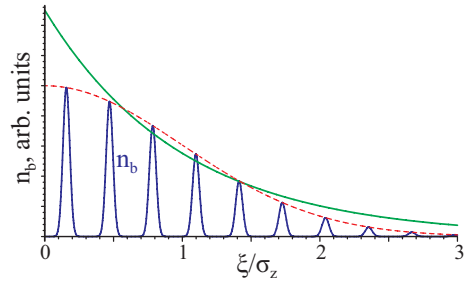}
\caption{A bunch train produced from a Gaussian beam [$n_b \propto \exp (-\xi^2/(2 \sigma_z^2))$] and its approximation by an exponential function [$\propto \exp(-\xi/\sigma_z)$].}\label{f6-exponent}
\end{figure}

The exponent of the bunch charge increase and the bunch-to-bunch distance are, apparently, most difficult to vary in experiments. The growth rate is determined by the beam shape before micro-bunching. This shape is unlikely to be exponential, but its rising part can be approximated by an exponential function (Fig.\,\ref{f6-exponent}), thus defining the product $(\gamma \cot \alpha)$ from \eref{e9}. The bunch-to-bunch distance and the bunch length are determined by the bunching technique, thus defining the parameter $\alpha$ and the required plasma density from \eref{e5}.

Preparing the bunch train in accordance with the above recipes is not sufficient for stable train propagation. The bunches must be in a radial equilibrium with the plasma wave, which is not straightforward to reach by injecting Gaussian bunches into the plasma \cite{PoP24-023119}. Early simulations of bunch trains indicate that the wave phase does not change much during bunch equilibration, if the initial driver emittance is sufficiently low \cite{PoP5-785,NIMA-410-461} or varies along the bunches \cite{NIMA-410-388}. However, that was demonstrated for constant-current non-equidistant bunches, so ways to practical realization of the stable equidistant bunch train is still an open question.

To conclude, a charged particle driver composed of short equidistant bunches can fully reside in focusing and decelerating phases of the wakefield, but the bunch charge must grow along the bunch train, and the plasma density must be somewhat higher than the density providing the exact resonance with the train.

\ack

This work was supported by Siberian Branch of the Russian Academy of Science, project No. 0305-2014-0016.


\section*{References}

\begin{thebibliography}{88}
\bibitem{NatPhot7-775}
    S. M. Hooker,
    Developments in laser-driven plasma accelerators.
    Nature Photon. \textbf{7}, 775 (2013).
\bibitem{PPCF56-084013}
	R.Assmann, et al. (AWAKE Collaboration),
	Proton-driven plasma wakefield acceleration: a path to the future of high-energy particle physics.
	Plasma Phys. Control. Fusion \textbf{56}, 084013 (2014).
\bibitem{UFN55-965}
    V.D.Shiltsev,
    High energy particle colliders: past 20 years, next 20 years and beyond.
    Physics -- Uspekhi \textbf{55}, 965 (2012).
\bibitem{RAST9-1}
    E.R.Colby and L.K.Len,
    Roadmap to the Future.
    Reviews of Accelerator Science and Technology \textbf{9}, 1 (2016).
\bibitem{RAST9-209}
    D.Schulte,
    Application of Advanced Accelerator Concepts for Colliders.
    Reviews of Accelerator Science and Technology \textbf{9}, 209 (2016).
\bibitem{RAST9-63}
    M.J.Hogan,
    Electron and Positron Beam-Driven Plasma Acceleration.
    Reviews of Accelerator Science and Technology \textbf{9}, 63 (2016).
\bibitem{RAST9-19}
    K.Nakajima,
    Laser-Driven Plasma Electron Acceleration and Radiation.
    Reviews of Accelerator Science and Technology \textbf{9}, 19 (2016).
\bibitem{RMP81-1229}
   E. Esarey, C. B. Schroeder, and W. P. Leemans,
   Physics of laser-driven plasma-based electron accelerators.
   Rev. Mod. Phys. \textbf{81}, 1229 (2009).
\bibitem{IEEE-PS24-252}
    E.Esarey, P.Sprangle, J.Krall, and A.Ting,
    Overview of plasma-based accelerator concepts.
    IEEE Trans. Plasma Sci. \textbf{24}, 252 (1996).
\bibitem{NIMA-829-3}
    A. Caldwell, et al. (AWAKE Collaboration),
    Path to AWAKE: Evolution of the concept.
    Nuclear Instr. Methods A \textbf{829}, 3 (2016).
\bibitem{PoP18-103101}
	A. Caldwell, K. V. Lotov,
	Plasma wakefield acceleration with a modulated proton bunch.
	Phys. Plasmas \textbf{18}, 103101 (2011).
\bibitem{RAST9-85}
    E.Adli and P.Muggli,
    Proton-Beam-Driven Plasma Acceleration.
    Reviews of Accelerator Science and Technology \textbf{9}, 85 (2016).
\bibitem{RuPAC16-303}
    I.Sheinman, A.Petrenko,
    High-energy micro-buncher based on the mm-wavelength dielectric structure.
    Proceedings of RuPAC2016 (St. Petersburg, Russia), p.303-306.
\bibitem{PRL104-255003}
	N.Kumar, A.Pukhov, and K.Lotov,
	Self-modulation instability of a long proton bunch in plasmas.
	Phys. Rev. Lett. \textbf{104}, 255003 (2010).
\bibitem{PoP22-103110}
	K. V. Lotov,
	Physics of beam self-modulation in plasma wakefield accelerators.
	Phys. Plasmas \textbf{22}, 103110 (2015).
\bibitem{AWAKE} AWAKE collaboration, to be published.
\bibitem{NIMA-829-76}
    E. Gschwendtner, et al. (AWAKE Collaboration),
    AWAKE, The Advanced Proton Driven Plasma Wakefield Acceleration Experiment at CERN.
    Nuclear Instr. Methods A \textbf{829}, 76 (2016).
\bibitem{Patric} P.Muggli, et al. (AWAKE Collaboration),
    AWAKE readiness for the study of the seeded self-modulation of a 400 GeV proton bunch.
    Plasma Physics and Controlled Fusion, to appear.
\bibitem{PoP22-123107}
    K. V. Lotov,
    Effect of beam emittance on self-modulation of long beams in plasma wakefield accelerators.
    Phys. Plasmas \textbf{22}, 123107 (2015).
\bibitem{PoP18-024501}
	K.V.Lotov,
	Controlled self-modulation of high energy beams in a plasma.
	Phys. Plasmas \textbf{18}, 024501 (2011).
\bibitem{PAcc20-171}
    P.Chen,
    A possible final focusing mechanism for linear colliders.
    Part. Accel. \textbf{20}, 171 (1987).
\bibitem{PoP20-083119}
    K.V.Lotov,
    Excitation of two-dimensional plasma wakefields by trains of equidistant particle bunches.
    Phys. Plasmas \textbf{20}, 083119 (2013).
\bibitem{AIP396-75}
    B.N.Breizman, P.Z.Chebotaev, A.M.Kudryavtsev, K.V.Lotov, and A.N.Skrinsky,
    Self-Focused Particle Beam Drivers for Plasma Wakefield Accelerators.
    AIP Conf. Proc. \textbf{396}, 75 (1997).
\bibitem{PoP24-023119}
    K.V. Lotov,
    Radial equilibrium of relativistic particle bunches in plasma wakefield accelerators.
    Phys. Plasmas \textbf{24}, 023119 (2017).
\bibitem{PoP5-785}
    K.V.Lotov,
    Simulation of ultrarelativistic beam dynamics in plasma wake-field accelerator.
    Phys. Plasmas \textbf{5}, 785 (1998).
\bibitem{NIMA-410-461}
    K.V.Lotov,
    Simulation of ultrarelativistic beam dynamics in the plasma wake-field accelerator.
    Nuclear Instr. Methods A \textbf{410}, 461 (1998).
\bibitem{NIMA-410-388}
    A.M.Kudryavtsev, K.V.Lotov, and A.N.Skrinsky,
    Plasma wake-field acceleration of high energies: Physics and perspectives.
    Nuclear Instr. Methods A \textbf{410}, 388 (1998).

\end{thebibliography}
\end{document}